\documentclass[%
reprint, 
amsmath,amssymb,
apl,
]{revtex4-1}

\usepackage{graphicx}
\usepackage{dcolumn}
\usepackage{bm}
\usepackage{color}
\usepackage{epstopdf}

\begin{document}


\title{Doppler-free frequency modulation spectroscopy of atomic erbium in a hollow cathode discharge cell}

\author{Henning Brammer}
\email{brammer@iap.uni-bonn.de}
\author{Jens Ulitzsch}%
\author{Riad Bourouis}%
\author{Martin Weitz}
\affiliation{
 Institut f\"ur Angewandte Physik, Universit\"at Bonn\\
 Wegelerstrasse 8, 53115 Bonn, Germany
}%

\date{\today}

\begin{abstract}
The erbium atomic system is a promising candidate for an atomic Bose-Einstein condensate of atoms with a non-vanishing orbital angular momentum ($L \neq 0$) of the electronic ground state. In this paper we report on the frequency stabilization of a blue external cavity diode laser system on the $400.91$ $nm$ laser cooling transition of atomic erbium. Doppler-free saturation spectroscopy is applied within a hollow cathode discharge tube to the corresponding electronic transition of several of the erbium isotopes. Using the technique of frequency modulation spectroscopy, a zero-crossing error signal is produced to lock the diode laser frequency on the atomic erbium resonance. The latter is taken as a reference laser to which a second main laser system, used for laser cooling of atomic erbium, is frequency stabilized.
\end{abstract}

\maketitle


\section{Introduction}

Experiments on laser cooling of atoms and atomic quantum gases require a high frequency stability of the cooling laser sources. For atoms with high melting points, as it is the case in several of the more recently explored non-alkali atomic systems with large magnetic dipole moment, as atomic chromium \cite{Chrom} or the rare earth elements erbium \cite{McClelland}, dysprosium \cite{Dysprosium} and thulium \cite{Thulium}, the use of gas reference cells with the investigated atom in sub-Doppler spectroscopy would require extremely high cell temperatures. Thus, stabilization techniques directly on the atomic beam used for loading of a magneto-optical trap have been investigated e.g. for atomic erbium \cite{McClelland}. An alternative is the use of hollow-cathode lamps for frequency stabilization of the cooling laser, and the use of sub-Doppler spectroscopy within such discharge cells has long been investigated in several publications (see e.g. \cite{Haensch}).

We here report the realization of Doppler-free frequency modulated saturation spectroscopy of atomic erbium in a hollow cathode discharge cell. A home built discharge cell was used to observe the $400.91$ $nm$ cooling line transition. The Doppler-free spectroscopy signal was used to provide reliable frequency stabilization of the cooling laser system within an rms linewidth of $1.3$ $MHz$, corresponding to some $4 \%$ of the $35.6$ $MHz$ natural linewidth of the strong blue erbium cooling transition.

The erbium atom in its $4f^{12} 6s^2$ $^3H_6$ ground state has a large orbital angular momentum of $L = 5$. All so far realized atomic quantum gases have a spherically symmetric S-ground state configuration (L = 0). With such a configuration, in far detuned laser fields with detuning above the upper state fine structure splitting, the trapping potential is determined by the scalar electronic polarizability \cite{Grimm}. In contrast, for atoms with a $L>0$ ground state, as is the case for atomic erbium, the trapping potential also for far detuned dissipation-less trapping laser fields becomes dependent on the internal atomic state (spin) \cite{Dubetsky}. This allows for state-dependent lattice manipulations, as are of interest in the context of quantum computing in optical lattices \cite{Jaksch,Jessen,Mandel}. Moreover, Raman transitions between different ground state spin projections become possible with e.g. Nd:YAG laser fields, which can allow for a Fourier-synthesis of in principle arbitrarily shaped lattice potentials using the technique of multiphoton lattices \cite{Ritt06,Salger07}. This has prospects for novel quantum phase transitions in e.g. strongly correlated frustrated lattice configurations \cite{Santos,Danshita,Buonsante}.

The present paper summarizes an approach to provide reliable frequency stabilization of a cooling laser system for atomic erbium. Recent work has shown that despite the complex level scheme of this atom, use of a single cooling laser frequency is sufficient for the operation of a magneto-optic trap \cite{McClelland}, as a starting point for further cooling and trapping techniques towards quantum degeneracy of this atom. Erbium has six stable bosonic isotopes and one stable fermionic isotope.

Section \ref{experiment} describes the experimental setup used for sub-Doppler spectroscopy and frequency locking based on the hollow cathode discharge lamp, and section \ref{results} gives experimental results. Finally, conclusions are presented in section \ref{conclusion}.

\section{Experimental Setup}\label{experiment}

\begin{figure}[b]
\includegraphics[width=0.47\textwidth]{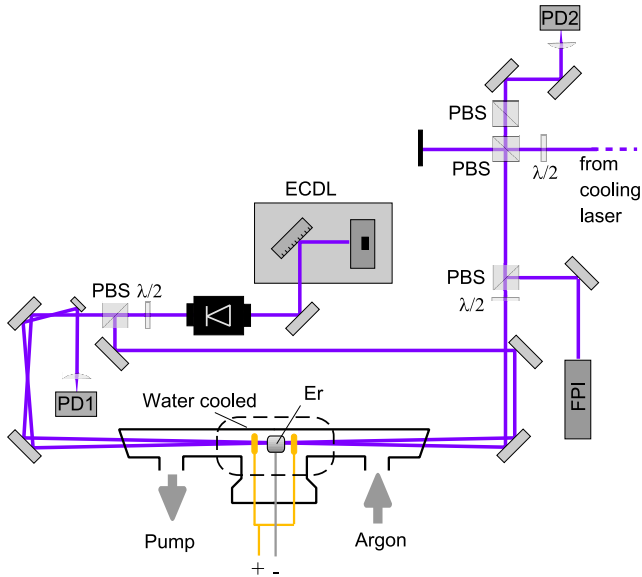}
\caption{\label{Abb:Skizze} Schematic diagram of the experimental setup. A laser beam derived from a grating stabilized diode is split into a pump and a probe beam. These counterpropagating beams cross in a erbium hollow cathode discharge cell. The probe beam is, after passage through the cell, imaged on photodiode $PD$1, from which the spectroscopy signal is derived. The pump beam is, after passage through the cell, reused in an offset frequency lock with the cooling laser frequency, and a corresponding beat signal is measured on $PD$2.}
\end{figure}

Erbium is a rare earth metal with a melting temperature of $1522$ $^\circ$C, and sufficient vapor pressure for spectroscopy in a regular gas cell would require cell temperatures well over $1000$ $^\circ$C. A hollow cathode cell, however, allows spectroscopy at much lower temperatures by sputtering \cite{Haensch,Smeets,Lister}.

We use a home built hollow cathode setup. The layout of the used optical setup along with a scheme of the spectroscopy cell is shown in Fig. \ref{Abb:Skizze}. The cell is made of a water-cooled glass tube of $20$ $cm$ length and $2.5$ $cm$ diameter. The basic geometry of the hollow cathode discharge follows a design used by the Stuttgart group for atomic chromium \cite{Pfau}. In this discharge, argon gas flows through the cell. The incoming flow through an entrance flange is carefully controlled with a needle valve and argon leaves the cell via a second flange connected to a rotary vacuum pump. The resulting cell pressure is measured with a vacuum gauge. The electrodes are mounted on one KF-flanged electrical feedthrough adapter and are placed in the middle of the cell. The anodes are made of isolated copper wire with dismantled ends formed to spirals. The cathode is formed by a center-drilled erbium slide of about $1$ $cm$ and $0.3$ $cm$ diameter and thickness respectively. The slide is mounted on a copper wire, which is as well as the cathodes isolated by ceramic tubes and Teflon strip. 

Both probe and counterpropagating pump laser beams can pass through the centered hole in the anodes and the cathode. The discharge is driven by a DC voltage supply, and a $10$ $k\Omega$ resistor was placed in series with the cell in order to stabilize the discharge current. The discharge in the hollow cathode lamp produces vaporised erbium atoms by sputtering, without the need of high temperatures. Typical operation parameters were an argon pressure of $1$ $mbar$, a discharge current of $70$ $mA$ and an operation voltage of $1.3$ $kV$. All measurements reported in this paper were recorded with these parameters unless otherwise noted. Experimentally, we found the operation of the discharge to be very critically dependent on the argon pressure.

Light for atomic spectroscopy of the $4f^{12}6s^{2}\,^{3}H_6 \rightarrow 4f^{12}(^{3}H_6)6s6p(^{1}p_1\!^o)(6, 1)_7\!^o$ atomic erbium transition near $400.91$ $nm$ was provided by a grating stabilized diode laser. We used a $Nichia$ $NDV$ $4313$ diode chip and a $3600$ $line/mm$ UV reflective holographic grating as an end reflector for the diode extended cavity. After passing an optical isolator, the emitted beam is split up into a pump and a probe beam respectively, which are directed in a counterpropagating geometry through the spectroscopy cell. The used optical powers for pump and probe beams were $2.4$ $mW$ and $0.6$ $mW$ respectively on an elliptical shaped beam with axis lengths of about $2$ and $1$ $mm$ respectively. After passing the cell, the probe beam is directed onto a fast photodiode.

With this setup we performed Doppler-free spectroscopy of the $400.91$ $nm$ erbium atomic resonance. For laser frequency stabilization, we use a frequency modulation (FM) technique to generate a dispersive error signal from the saturation spectroscopy signal \cite{bjorklund}. The diode laser injection current is modulated with a frequency of 20 MHz, which results into a modulation of the diode laser frequency. The photodiode (PD1) signal is amplified and mixed with part of the radiofrequency signal used for diode laser modulation using standard techniques, which for a proper choice of the relative phase between the photodiode signal and the reference local oscillator signal results in a dispersive error signal.

By feeding the integrated FM spectroscopy signal back to the piezo-mounted grating of the diode laser, the frequency of the diode laser system can be stabilized onto the Doppler-free erbium atomic resonance. In further measurements, this laser is used as a reference for a frequency offset locking of the cooling laser. By overlapping the pump beam after transmission through the spectroscopy cell with a beam derived from a frequency doubled Ti:sapphire laser ($1.5$ $W$ output power at $401$ $nm$, this optical source will be used for laser cooling of atomic erbium) on a photodiode (PD2), as shown in Fig. \ref{Abb:Skizze}, a beat note between the two laser frequencies is obtained. With the help of a frequency to voltage converter an error signal used for a frequency offset locking of the cooling laser system relatively to the reference laser frequency is obtained.

\section{Results and discussion} \label{results}


\begin{figure}[t]
\includegraphics[width=0.47\textwidth]{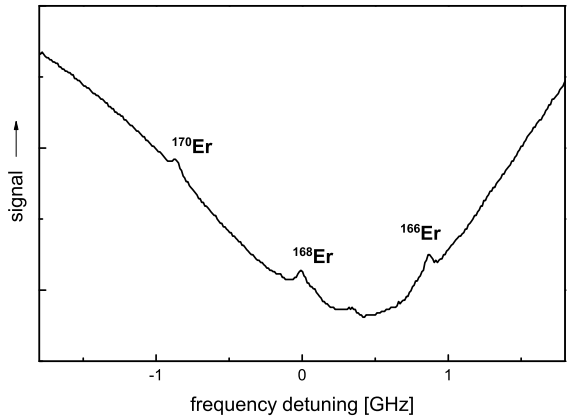}
\caption{\label{Abb:Saturation} Saturation spectroscopy signal of the $400.91$ $nm$ erbium line. The lamb dips of the corresponding spectral lines of the three bosonic isotopes  $^{166}$Er, $^{168}$Er and $^{170}$Er and some of the hyperfine components of the fermionic isotope $^{167}$Er can be seen.}
\end{figure}

Typical observed saturation spectroscopy signals of the $400.91$ $nm$ line of atomic erbium are shown in Fig.\ref{Abb:Saturation}. The frequency axis in this plot, as well as for the other shown spectra, was calibrated using a Fabry-Perot interferometer. The plot shows a Doppler-broadened background along with three Doppler-free saturation peaks, which are attributed to the Doppler-free resonances of the three bosonic isotopes with the highest abundance, $^{166}$Er (33.4\%), $^{168}$Er (27.1\%) and $^{170}$Er (14.9\%), and the fermionic isotope $^{167}$Er (22.9\%). While the bosonic isotopes have no hyperfine structure (nuclear spin $I=0$), the fermionic isotope with its 7/2 $\hbar$ nuclear spin splits into several peaks. The spectral splitting between the bosonic $^{170}$Er and $^{168}$Er isotopes was measured to be (855$\pm$10) $M\!H\!z$ (and (859$\pm$10) $M\!H\!z$ between $^{168}$Er and $^{166}$Er), in good agreement with our fluorescence measurements realized so far by the use of an erbium atomic beam, as well as with older fluorescence measurements \cite{ross}. The broad Doppler-sensitive absorption profile visible in Fig.\ref{Abb:Saturation} is caused by contributions of the Doppler-broadened resonances of several of the erbium isotopes. To estimate the temperature of the atomic erbium gas in the hollow cathode discharge cell, we have fitted the sum of Doppler-broadened gaussian line profiles for the three dominant bosonic isotope peaks to the observed line profile. This procedure results in a temperature value of (470$\pm$90)$^\circ$C, which one may take as an estimate for the temperature of the gas. This value is approximately three times smaller than the melting point of erbium.

\begin{figure}[t]
\includegraphics[width=0.47\textwidth]{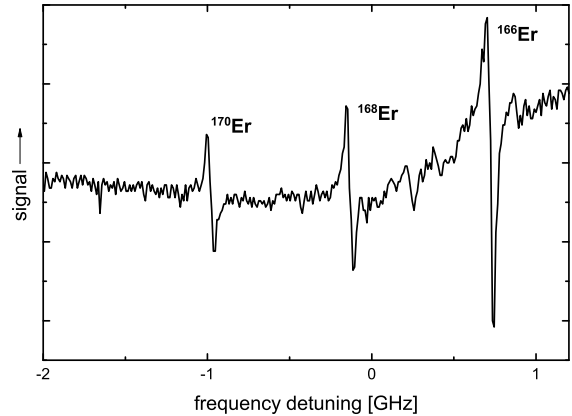}
\caption{\label{Abb:FM} Demodulated FM spectroscopy signal of atomic erbium versus laser frequency. Due to the dispersive shapes of the signal around the resonances, this signal can be directly used as an error signal for stabilization of the laser frequency.}
\end{figure}

The observed linewidth of the transitions (around $100$ $M\!H\!z$) is roughly a factor 2-3 bigger than the natural linewidth of the corrsponding erbium resonance. As the observed linewidth was not very strongly dependent on the pressure of the argon gas, we attribute the additional broadening not to be dominated by pressure broadening, but rather to be mainly determined by contributions from saturation broadening and the Stark effect, which in discharge lamps due to different electric field values within the dc discharge results in a spectral broadening. \cite{Lister}.

\begin{figure}[t,b]
\includegraphics[width=0.47\textwidth]{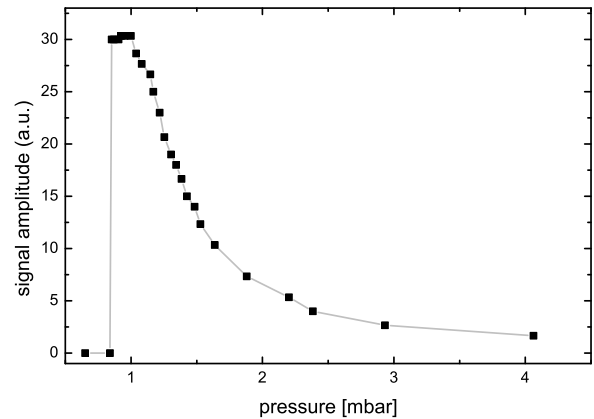}
\caption{\label{Abb:Druck} The graph shows the amplitude of the FM-spectroscopy signal versus the gas pressure in the hollow cathode discharge cell. A large signal amplitude and correspondingly also a high signal to noise ratio is observed only for a restricted pressure range around $1$ $mbar$. The argon pressure is  controlled with the help of a needle valve.}
\end{figure}

Fig. \ref{Abb:FM} shows a typical signal obtained using the FM spectroscopy technique. The Doppler-sensitive background is here reduced, and the shown spectrally sharp Doppler-free resonances (three of them can be clearly identified to transitions of the corresponding bosonic isotopes, as shown in the figure) have dispersive character. Fig. \ref{Abb:Druck} shows the amplitude of the observed signal for the case of the $^{168}$Er line with large natural abundance versus the argon pressure in the discharge. One clearly observes a large decrease of the signal amplitude for larger pressure values. The visible sharp drop at pressure values below $0.9$ $mbar$ is due to the issue that below this pressure value the discharge could not be reliably operated for the used parameters. The usual operation pressure for the discharge was, as described previously, an argon pressure of $1$ $mbar$, for which the obtained signal amplitude reaches its maximum value. As we did not observe a clear increase of the signal linewidth with pressure, we suspect that the signal drop with pressure is not due to pressure broadening in the discharge, but rather to the efficiency in the production of free erbium atoms in the spatial region experienced by the optical spectroscopy beams.

By feeding the integrated and amplified FM spectroscopy signal back to the piezo steering of the grating of the external cavity diode laser system, the laser frequency can be stabilized onto the atomic transition frequencies of any of the three bosonic erbium isotopes. The measured frequency stability of the diode laser relatively to the atomic resonance in this servo loop was $1.3$ $M\!H\!z$ for the case of the $^{166}$Er resonance.

We have furthermore investigated the frequency offset locking of the doubled Ti:sapphire laser, which will be used for laser cooling of erbium atoms in our experiment, relatively to the diode laser frequency. The frequency stability of the offset lock was measured to be $0.9$ $M\!H\!z$, resulting in an overall frequency stability of the cooling laser relatively to the atomic resonance of about $1.6$ $M\!H\!z$, which is about a factor 20 below the $35.6$ $M\!H\!z$ natural linewidth. We thus expect that the obtained frequency stability is sufficient for laser cooling of atomic erbium on the used blue cooling resonance. By applying an offset voltage on the beatlock, the frequency of the cooling laser can be scanned over a wide range within the erbium spectrum, and e.g. also onto one of the transitions of the fermionic $^{167}$Er atomic isotope.

\section{Conclusion} \label{conclusion}

We have demonstrated Doppler-free frequency modulated spectroscopy of atomic erbium within a hollow cathode discharge cell. Using saturation spectroscopy, we could resolve the Lamb dips of the different isotopes in the cell of the blue $400.91$ $nm$ erbium transition. For the sake of absolute frequency stabilization and increased signal to noise ratio, a zero crossing error signal has been produced using an FM-modulation technique within the hollow cathode spectroscopy. A frequency-doubled Ti:Sapphire laser, which will be used for laser cooling of atomic erbium, has been offset locked relative to the diode hollow cathode reference laser system.

In conclusion, frequency modulated Doppler-free spectroscopy on hollow cathode discharges allows for reliable frequency stabilization onto non-volatile atomic species, as is of interest in the context of laser cooling of such atomic species.


\bibliography{lit}

\end{document}